\documentclass[%
 reprint,
superscriptaddress,
 amsmath,amssymb,
]{revtex4-1}

\usepackage{bbold}
\usepackage{mathptmx}
\usepackage{subfig}
\usepackage{psfrag,graphicx}
\usepackage{dcolumn}
\usepackage{amsmath,amssymb}
\usepackage{bm}
\usepackage{color}
\usepackage{latexsym}
\usepackage{epstopdf}
\usepackage{color}
\usepackage[english]{babel}
\usepackage{latexsym}
\usepackage{psfrag,graphicx}
\usepackage{amsmath}
\usepackage{amssymb}
\usepackage{amsfonts}
\usepackage{bm}
\usepackage{natbib}
\usepackage{epstopdf}
\usepackage{appendix}
\usepackage{rotating}
\usepackage[english]{babel}
\usepackage{aeguill}
\usepackage{ulem}
\usepackage[justification=justified]{caption}

\definecolor{mygrey}{gray}{0.35}
\definecolor{myblue}{rgb}{0.2,0.2,0.8}
\definecolor{myzard}{cmyk}{0,0,0.05,0}
\definecolor{mywhite}{rgb}{1,1,1}
\definecolor{mywhite}{rgb}{1,1,1}
\definecolor{myred}{rgb}{1,0.,0.3}

\usepackage[colorlinks=true,citecolor=myblue,linkcolor=myblue]{hyperref}

\def\ba{\begin{align}}
\def\enda{\end{align}}
\def\bi{\begin{itemize}}
\def\ei{\end{itemize}}

\def\be{\begin{equation}}
\def\ee{\end{equation}}
\def\bea{\begin{eqnarray}}
\def\eea{\end{eqnarray}}
\def\bse{\begin{subequations}}
\def\ese{\end{subequations}}

\newcommand{\ket}[1]{|{#1}\rangle}                       % ket
\newcommand{\bra}[1]{\langle {#1}|}                      % bra
\newcommand{\average}[1]{\langle {#1} \rangle}           % media < >

\newcommand{\ketbra}[2]{\left\vert#1\right\rangle\left\langle#2\right\vert}

\newcommand{\Ignore}[1]{ }

\def\i{\text{i}}

\begin{document}

\preprint{APS/123-QED}

\title{Spin-spin coupling-based quantum and classical phase transitions in two-impurity spin-boson models}

%\Ignore{
\author{R. Grimaudo}
\address{Dipartimento di Fisica e Chimica ``Emilio Segr\`{e}",
Universit\`{a} degli Studi di Palermo, viale delle Scienze, Ed. 18, I-90128, Palermo, Italy}

\author{A. Messina}
%\address{INFN, Sezione di Catania, I-95123 Catania, Italy}
\address{ Dipartimento di Matematica ed Informatica, Universit\`a degli Studi di Palermo, Via Archirafi 34, I-90123 Palermo, Italy}

\author{H. Nakazato}
\address{Department of Physics, Waseda University, Tokyo 169-8555, Japan}
\address{Institute for Advanced Theoretical and Experimental Physics, Waseda University, Tokyo 169-8555, Japan}

\author{A. Sergi}
\address{Dipartimento di Scienze Matematiche e Informatiche, Scienze Fisiche e Scienze della Terra, Universit\`{a} degli Studi di Messina, Viale F. Stagno d'Alcontres 31, 98166 Messina, Italy}
\address{Istituto Nazionale di Fisica Nucleare, Sez. di Catania, 95123 Catania, Italy}
\address{Institute of Systems Science, Durban University of Technology, P.O. Box 1334, Durban 4000, South Africa}

\author{D. Valenti}
\address{Dipartimento di Fisica e Chimica ``Emilio Segr\`{e}",
Universit\`{a} degli Studi di Palermo, viale delle Scienze, Ed. 18, I-90128, Palermo, Italy}
%}

\date{\today}

\begin{abstract}
 
The class of two-interacting-impurity spin-boson models with vanishing transverse fields on the spin-pair is studied.
The model can be exactly mapped into two independent standard single-impurity spin-boson models where the role of the tunnelling parameter is played by the spin-spin coupling.
The dynamics of the magnetization is analysed for different levels of (an)isotropy.
Further, the existence of a decoherence-free subspace as well as of both classical and quantum (first-order and Kosterlitz-Thouless type) phase transitions, in the Omhic regime,  is brought to light.
 
\end{abstract}



\maketitle

The dynamics of any open quantum system is profoundly influenced by its surrounding environment which is at the origin of decoherence and/or dissipation manifestation \cite{BP02}.
The former effect plays a leading role in determining the transition from quantum to classical behavior.
In the last decades, it has attracted much attention mostly in the field of quantum state manipulation and quantum computation \cite{NC10}.

An important model exhibiting quantum dissipation is the so called single-impurity spin-boson model (SISBM) which describes a single spin-1/2 coupled to a bosonic quantum bath \cite{Leggett87}.
The SISBM has been thoroughly studied in wide regions of the parameter space with diverse methods and techniques since the 1980s \cite{Leggett87, LeHur07, Vojta05, Bulla03}.
%The SISBM 2,3 is a popular starting point for investigations about the dissipative dynamics of a noisy two-level system or qubit.
It encapsulates effects stemming from quantum decoherence, dissipation, and relaxation on the otherwise coherent spin evolution \cite{Leggett87}.
Furthermore, the model exhibits a nontrivial ground-state behavior, since it displays a quantum phase transition as a function of system-bath coupling strength \cite{LeHur08, Nazir12}, attributed to zero-point rather than thermal fluctuations within the bath \cite{Vojta03, Rossini21, Carollo20}.
Applications are numerous, ranging from quantum optics to quantum information and computation \cite{Kundu21, Dunnett21, Lemmer18, Lerma19, Leppakangas18, Puebla19, Wenderoth21, Magazzu18, DeFilippis20, Wang19, Wang20, Lemmer18, Shen21, Aurell20, Miessen21, Villase20, Pino18, Magazzu18nat, Lambert19, Casanova18}.

The interest towards decoherence and dissipation as well as quantum phase transitions (QPTs) in two-impurity spin-boson models (TISBMs) with competing interactions has remarkably grown in the last two decades \cite{Dolgitzer21, Wang21, Zhou18, Nagele10, Storcz05, Garst04, McCutcheon10, Bonart13, Orth10, Zheng15, Winter14, Nagele08, Thorwart01, Storcz03}.
The TISBM is currently under attention to determine the existence of critical points and then the presence of quantum and/or classical phase transitions \cite{Zhou18, Wang21}.
On the basis of numerical approaches, different results have been proposed; however, until now, a univocal response is missing \cite{Zhou18, Wang21}.
Moreover, a remarkable question to be addressed is whether, in presence of the impurity-impurity coupling, the transition is of the Kosterlitz-Thouless (K-T) type \cite{Zhou18, Wang21}.

In this work we study the class of TISBMs useful for describing bi-nuclear units \cite{Calvo11, Napolitano08}, where the transverse ($x$) field on the spins is absent and a non-isotropic spin-spin Heisenberg interaction is considered.
We show that the dynamical problem can be exactly and analytically reduced to that of two independent SISBMs, wherein the role of the transverse field is effectively played by the two-spin coupling(s).
First, we bring to light the existence of a decoherence-free subspace, charcterized then by a dissipationless spin-dynamics. 
Further, basing on the results previously obtained for the SISBM in the Ohmic case \cite{Leggett87, LeHur08, Nazir12}, we derive the behaviour of the magnetization of the system as well as the presence of both classical and QPTs.
In particular, two types of QPT are present: a first-order QPT (due to a level crossing) and a K-T QPT.

\textit{Model.}
Consider the following model (in units of $\hbar$):
\begin{equation} \label{Hamiltonian}
\begin{aligned}
{H} = &
{\Omega_{1} \over 2}\hat{\sigma}_{1}^{z}+{\Omega_{2} \over 2}\hat{\sigma}_{2}^{z} + \sum_{j=1}^N  \omega_j ~ \hat{a}_j^\dagger \hat{a}_j -\\
&{\gamma_{x} \over 2}\hat{\sigma}_{1}^{x}\hat{\sigma}_{2}^{x} - {\gamma_{y} \over 2}\hat{\sigma}_{1}^{y}\hat{\sigma}_{2}^{y}-\gamma_{z}\hat{\sigma}_{1}^{z}\hat{\sigma}_{2}^{z}+
\sum_{k=1}^2\sum_{j=1}^N {c_{kj} \over 2} \left( \hat{a}_j^\dagger + \hat{a}_j \right) \hat{\sigma}_k^z,
\end{aligned}
\end{equation}
which describes two interacting spin-1/2's subject to local longitudinal ($z$) fields and coupled to a common bath of quantum harmonic oscillators.
$\Omega_i$ and $\omega_j$ are the characteristic frequencies of the $i$-th spin and the $j$-th mode, respectively.
$\hat{\sigma}_{k}^{l}$ ($k=1,2$, $l=x,y,z$) are the Pauli operators of the spins, while $a_j$ and $a_j^\dagger$ are the annihilation and creation boson operators of each field mode.

Thanks to the existence of the constant of motion $\hat{\sigma}_{1}^{z}\hat{\sigma}_{2}^{z}$, the model can be unitarily transformed into $\widetilde{H} = \widetilde{H}_a \oplus \widetilde{H}_b$,% (see Supplemental Material),
%\begin{equation}
%    \widetilde{H} = \widetilde{H}_a \oplus \widetilde{H}_b,
%\end{equation}
with
\begin{equation}
\begin{aligned}
{\widetilde{H}}_{a/b} = &
{\Omega_{a/b} \over 2} \hat{\sigma}_{a/b}^{z} -
{\gamma_{a/b} \over 2} \hat{\sigma}_{a/b}^{x} \mp \gamma_{z} \hat{\mathbb{1}}_{a/b} + \\
&\sum_{j=1}^N   \omega_j ~ \hat{a}_j^\dagger \hat{a}_j +
\sum_{j=1}^N {c_j^{a/b} \over 2}  \left( \hat{a}_j^\dagger + \hat{a}_j \right) \hat{\sigma}_{a/b}^z \label{eff ham 1},
\Ignore{
\\
{\widetilde{H}}_b =&
{\Omega_b \over 2} \hat{\sigma}_b^{z} -
\sum_{j=1}^N   \omega_j ~ \hat{a}_j^\dagger \hat{a}_j +
\sum_{j=1}^N {c_j^b \over 2}  \left( \hat{a}_j^\dagger + \hat{a}_j \right) \hat{\sigma}_b^z, \label{eff ham 2}
}
\end{aligned}
\end{equation}
where $\Omega_{a/b} = \Omega_1 \pm \Omega_2$, $\gamma_{b/a} = \gamma_{x} \pm \gamma_{y}$, and $c_j^{a/b} = c_{1j} \pm c_{2j}$.
$\widetilde{H}_a$ and $\widetilde{H}_b$ are effective Hamiltonians governing the dynamics of the two-spin-boson system (TSBS) within each dynamically invariant subspace.% related to one of the two eigenvalues of $\hat{\sigma}_1^z\hat{\sigma}_2^z$.

The two independent subdynamics are equivalent to two effective SISBMs: I) the coupling between the two true spins provides the effective transverse magnetic field ($\gamma_a$ and $\gamma_b$); II) the longitudinal field results from precise combinations ($\Omega_a$ and $\Omega_b$) of the two fields applied to the actual spin-1/2's; III) the coupling with the quantum oscillator bath is mediated by appropriate combinations ($c_j^a$ and $c_j^b$) of the coupling parameters of the two spin-1/2's with each boson mode.
Therefore, all the results obtained  for the SISBM can be applied to each subdynamics and exploited to get information about the TSBS dynamics.
%the unitary mapping between the hamiltonian models 1 and 2, bridges the two physcal scenarions in an effectively expoitable way.

\textit{Decoherence-Free Subspace.}
%First of all, it is worth noticing that the subspace $b$ related to the eigenvalue -1 of $\hat{\sigma}_1^z\hat{\sigma}_2^z$ (spanned by $\{\ket{+-},\ket{-+}\}$) presents the following peculiar dynamical property.
If $c_j^b=0$ the subspace $b$ is a decoherence-free subspace.
It means that, although the actual spins interact with the bath, they experience a dissipationless dynamics within such a subspace as if the bath were absent.
Therefore, the initial or produced entanglement between the spins would not degrade, despite the presence of the spin-bath coupling term. 

%It is particularly interesting that this dynamical feature holds if the two spins are characterized by the same coupling with each boson field mode, a relatively simple condition to fulfil through a structured environment.
This circumstance is of remarkable importance in quantum computation, where controlling the dissipative spin-spin-boson dynamics in nonequilibrium conditions, e.g., in the presence of time-dependent external fields, is crucial \cite{Orth10, Thorwart01}.
We emphasize that the analytical treatment employed to unitary transform the TISBM is not affected by any time-dependence of the Hamiltonian parameters.% (see Supplemental Material).
In this way, appropriate time variations of the local fields %$\Omega_1(t)$ and $\Omega_2(t)$
and/or the coupling parameters %$\gamma_x(t)$ and $\gamma_y(t)$
can be engineered in order to generate unperturbed quantum gates acting on the two spins.

\textit{Conditions.}
Consider the bath in a thermal state and the two-spin system prepared in the state $\rho(0)=\ketbra{++}{++}$, which is mapped to the single-spin state $\rho_a(0)=\ket{+}_a\bra{+}$.
In this instance, the dynamics is entirely restricted to the subspace $a$ (spanned by $\{\ket{++},\ket{--}\}$) and the mean value of the total magnetization $\average{\hat{\Sigma}^z} \equiv \average{\hat{\sigma}_1^z} + \average{\hat{\sigma}_2^z}$, as well as the mean value of the single magnetizations of the two spins, can be easily obtained from $\average{\hat{\sigma}_a^z}$:% (see Supplemental Material):
%Indeed, for the case under scrutiny, we have
\begin{equation}
    \text{Tr}\{\hat{\rho}_a(t)\hat{\sigma}_a^z\} =  \average{\hat{\sigma}_a^z} = \average{\hat{\sigma}_1^z} = \average{\hat{\sigma}_2^z} = {\average{\hat{\Sigma}^z} \over 2 }.
\end{equation}

%Before starting, it is important to define the bath spectral density function $J(\omega)= \pi \sum_j {(c_j^+)^2} \delta(\omega - \omega_j)$ (we stress that the effective coupling(s) $c_j^+$ enters the definition of the function).
In the asymptotic low-temperature limit the bath spectral density function $J(\omega)= \pi \sum_j {(c_j^a)^2} \delta(\omega - \omega_j)$ is determined by the low-energy part of the spectrum and its standard parametrization is
\begin{equation}
    J(\omega) = 2 \pi \alpha \omega_c^{1-s} \omega^s,
    \qquad
    0 < \omega < \omega_c,
    \quad
    s > -1,
\end{equation}
where $\omega_c$ is a cut-off frequency and $\alpha$ is the dimensionless parameter accounting for the dissipation strength \cite{Leggett87}.
The spectral exponent $s$ defines three regimes: Ohmic ($s=1$), sub-Ohmic ($s<1$) and super-Ohmic ($s>1$).

\textit{Ohmic Regime.}
The ohmic case ($s=1$) presents a large variety of different behaviours depending on the region of the parameter space taken into account \cite{Leggett87}.
First, consider the case $\Omega_1=\Omega_2=0$.
We underline that all the following results are valid under two conditions \cite{Leggett87}: 1) $\Omega_{a/b}$, $\gamma_{b/a}$ and $k_BT$ are small compared to the bath cut-off frequency $\omega_c$; 2) the `interesting' times are large compared to $\omega_c^{-1}$ \cite{Leggett87}.

For $\alpha=1/2$ the two-spin magnetization reads \cite{Leggett87}
\begin{equation} \label{P s1 a1/2}
    \average{\hat{\Sigma}^z(t)} = 2 \exp \left\{ -{\pi \over 2 } {\gamma_a^2 \over \omega_c} t \right\}.
\end{equation}
This result is valid at all temperatures ($T=0$ and $T \neq 0$) compatible with the condition $k_BT \ll  \omega_c$ \cite{Leggett87}.
The exponential decaying rate of the two-spin probability depends on the ratio $\gamma_a^2 / \omega_c$, meaning that the characteristic time-scale of the system is determined by the spin-spin coupling parameter.
Precisely, it depends on the difference $\gamma_x - \gamma_y$, so that: i) in case of isotropy ($\gamma_x=\gamma_y$) the system tends to remain in its initial state; ii) a slight difference between the two coupling parameters, instead, causes an exponential decay of the magnetization(s) towards the equilibrium value.

For $\alpha<1$ ($\neq 1/2$) two cases can be considered: $k_B T \gtrsim \widetilde{\gamma}_a$ and $k_B T \lesssim \widetilde{\gamma}_a$, with \cite{Leggett87}
\begin{equation}
    \widetilde{\gamma}_a = \gamma_a \left( {\gamma_a \over  \omega_c} \right)^{\alpha / 1-\alpha}.
\end{equation}
In the first case the magnetization results to be \cite{Leggett87}
\begin{subequations} \label{Exp Relax}
  \begin{align}
      \average{\hat{\Sigma}^z(t)} &= 2\exp \{ -t / \tau \}, \\
      %\tau^{-1} &= {\sqrt{\pi} \over 2} { \Gamma(\alpha) \over \Gamma(\alpha+1/2)} {\widetilde{\gamma}_a^2 \over  k_B T} \left( \pi k_B T \over \widetilde{\gamma}_a \right)^{2\alpha} \\ \nonumber
      \tau^{-1} &= {\sqrt{\pi} \over 2} { \Gamma(\alpha) \over \Gamma(\alpha+1/2)} {\gamma_a^2 \over \omega_c} \left( \pi k_B T \over  \omega_c \right)^{2\alpha-1},
  \end{align}
\end{subequations}
where $\Gamma$ is the gamma function.
The previous expression describes an exponential relaxation with a rate $\propto T^{2\alpha-1}$.
In the second case: i) if $1/2<\alpha<1$ the time behavior is most likely an incoherent relaxation with an $\alpha$-dependent rate of order $\widetilde{\gamma}_a^{-1}$; ii) if $0<\alpha<1/2$ the system exhibits damped incoherent oscillations \cite{Leggett87}.
These results show that, depending on the ratio of the spin-spin energy coupling to the thermal energy, different dynamics arise.
Therefore, the spin-spin interaction, besides the decaying rate, sets the limit temperature dividing the two dynamical regions.
Physical systems characterized by different couplings exhibit thus a different critical temperature and/or different behaviours at the same temperature.
Three scenarios can be considered.
In nuclear magnetic resonance, the spin-spin coupling typically ranges from 10 Hz to 300 Hz, depending on the molecule \cite{Vandersypen05}.
For microwave-driven trapped ions the interaction strength can reach the kHz range \cite{Weidt16}.
Rydberg atoms and ions, due to the huge electric-dipole moments of the Rydberg states, are characterized by an effective spin-spin coupling which can reach a few MHz \cite{Gaetan09, Urban09}.
In the three cases the critical temperature $T_c=\widetilde{\gamma}_a/k_B$ separating the two dynamical regimes results $T_c \approx 0.1-1 ~ nK$, $T_c \approx 10 ~ nK$, $T_c \approx 10 ~ \mu K$, respectively.

When $\alpha>1$: i) if $T \neq 0$, the two-spin dynamics consists in the same exponential relaxation written in Eq. \eqref{Exp Relax}, characterized by a rate $\propto T^{2\alpha-1}$ \cite{Leggett87}; ii) for $T=0$, instead, the two-spin system experiences the localization regime, that is, it is frozen in its initial condition \cite{Leggett87}.

It is worth noticing that all the previous dynamical behaviours rely only on internal parameter characterizing the physical system: the spin-spin coupling, the spin(s)-bath coupling and the bath cut-off frequency.
This aspect suggests a sort of self-organization of the system and an auto-determination of the system dynamics.

%\textit{Biased Case.}
An appropriate non-vanishing bias ($\Omega_a \ll \omega_c$), large compared to the renormalized tunneling frequency ($ \Omega \gg \widetilde{\gamma}_a$), makes the system to relax from the upper to the lower state, even at zero temperature \cite{Leggett87}.
Thus, the physical effect of a sufficient bias is that to suppress the coherent oscillations shown, in some cases, by the unbiased system.

\textit{Mixing Subspaces}.
All the previous results can be even applied on the subdynamics $b$, with $\ket{+-}$ as initial state of the two spins and $\Omega_a$ and $\gamma_a$ replaced with $\Omega_b$ and $\gamma_b$, respectively.
Of course, the Hamiltonian parameters are constrained to fulfill $\gamma_b \ll \omega_c$ and $\Omega_a \ll \omega_c$.
In this case net magnetization vanishes, $\average{\hat{\Sigma}^z}=0$, since $\average{\hat{\sigma}_1^z}=-\average{\hat{\sigma}_2^z}$. 

\begin{figure}[b!!] 
\begin{center}
{\includegraphics[width=0.45\textwidth]{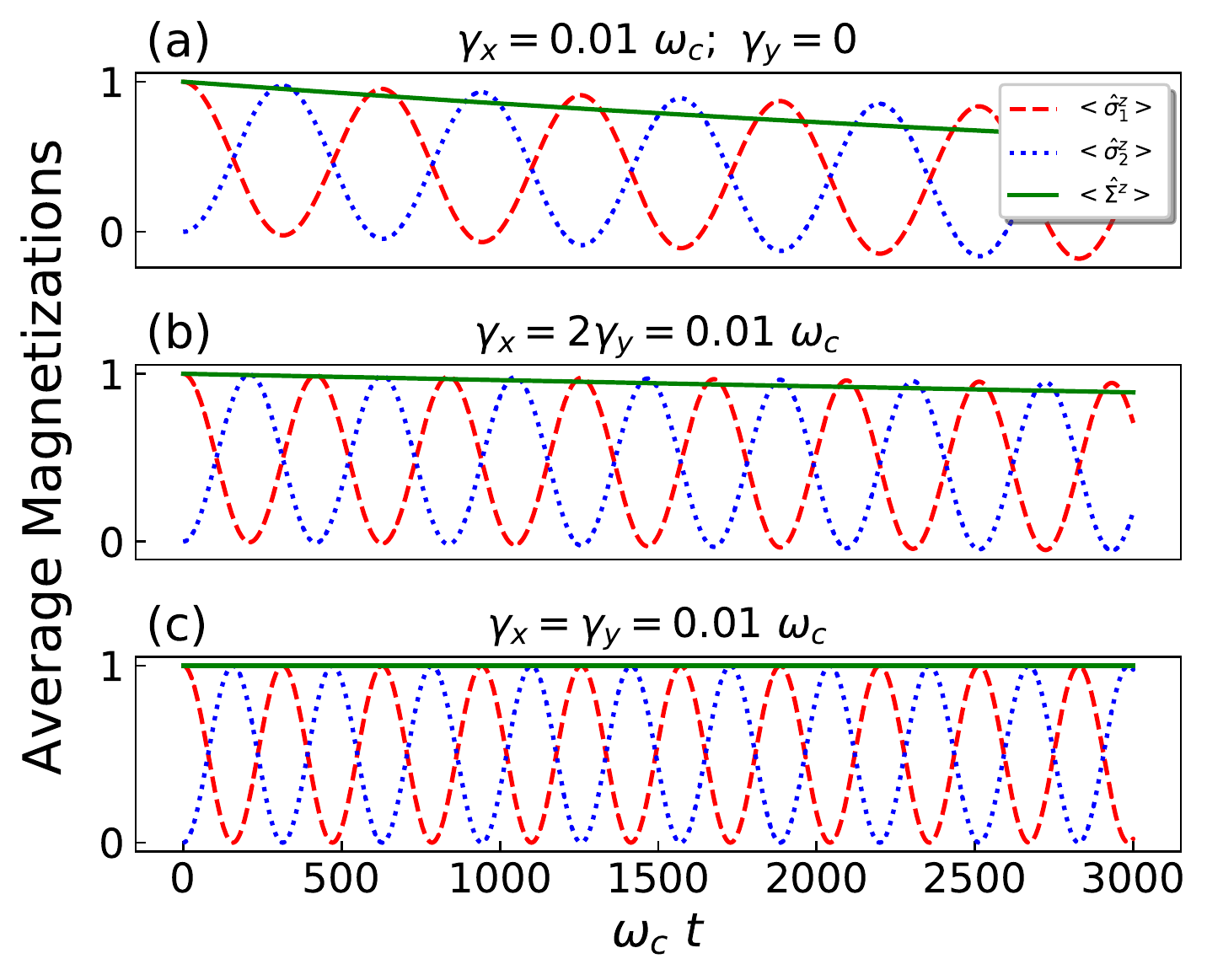}}
\captionsetup{justification=raggedright,format=plain,skip=4pt}%
\caption{Time behaviour of both the single-spin and the net magnetizations
%$\average{\hat{\sigma}_1^z}$ (red dashed line), $\average{\hat{\sigma}_2^z}$ (blue dotted line) and $\average{\hat{\Sigma}^z}$ (green solid line)
for $s=1$, $\alpha_a=1/2$, and three levels of isotropy: (a) minimum, (b) intermediate, (c) maximum.
The two spins start from $(\ket{++}+\ket{+-})/\sqrt{2}$, while the bath from the thermal state.}
%Three cases are considered: (a) maximum anisotropy: $\gamma_x = 0.01  \omega_c$, $\gamma_y=0$; (b) intermediate anisotropy: $\gamma_x = 2 \gamma_y = 0.01  \omega_c$; (c) isotropy: $\gamma_x = \gamma_y = 0.01  \omega_c$.}
\label{fig: magnetizations}
\end{center}
\end{figure}

If we consider the initial condition $(\ket{++}+\ket{+-}) / \sqrt{2}$, both subspaces are involved.
In this circumstance, we have to indipendently solve the dynamics in each subspace and `merge' the results.
It must be taken into account that the two subdynamics are characterized by different (effective) couplings with the bath and then by different $\alpha$s, say $\alpha_a$ and $\alpha_b$.
Consider the following case: $s=1$, $\alpha_a=1/2$ and $\alpha_b=0$.
The latter condition stems from $c_{1j}=c_{2j}, ~ \forall j$, so that $c_j^b=0$.

The time behaviour of the net magnetization $\average{\hat{\Sigma}^z}$ is determined by the time evolution in the subspace $a$ since no contribution stems from the subspace $b$.
Therefore, this time, the value of the magnetization is half times that obtained when the system is initially prepared in $\ket{++}$ [Eq. \eqref{P s1 a1/2}].%, then $\average{\hat{\Sigma}^z}=\average{\hat{\sigma}_a^z}$.

Rather, a relevant difference is found for $\average{\hat{\sigma}_1^z}$ and $\average{\hat{\sigma}_2^z}$.
Previously indeed we had $\average{\hat{\sigma}_1^z}=\average{\hat{\sigma}_2^z}=\average{\hat{\sigma}_a^z}$ (Eq. \eqref{P s1 a1/2}).
Now, the time-behaviour of each spin magnetization reads% (see Supplemental Material)
\begin{equation}
    \average{\hat{\sigma}_{1,2}^z} = {\average{\hat{\sigma}_{a}^z} \pm \average{\hat{\sigma}_b^z} \over 2} =  {e^{-{\pi \over 2} {\gamma_a^2 \over \omega_c} t} \pm \cos \left( \gamma_b t \right) \over 2}.
\end{equation}
The \textit{cosin}-term stems from the exact solution of the deterministic dynamics in the subspace $b$.

From Figs. \ref{fig: magnetizations}(a) and \ref{fig: magnetizations}(b) we see that the level of anisotropy influences both the frequency of the single-spin oscillations and the decaying rate of the net magnetization.
In case of isotropy ($\gamma_x=\gamma_y$), instead, the two spins exhibit dissipationless oscillations and the net magnetization is constant.
This circumstance is related to the vanishing value of $\gamma_a$ which rules the exponential relaxation (playing the role of the effective transverse field in the subdynamics $a$).
Therefore, by studying both the oscillation frequency of each spin magnetization and the exponential decaying rate of the net magnetization, the coupling parameters $\gamma_x$ and $\gamma_y$ and then the level of (an)isotropy of the two-spin system can be estimated.

\textit{QPT.}
Depending on the parameter-space region, the ground state (GS) of the two-spin system belongs to either the $a$ or $b$ space and coincides with the GS of the fictitious spin-qubit $a$ or $b$.
In this way, we can write the two possible GSs and the related ground energies of the TISBM on the basis of the expressions obtained for the SISBM.
The ansatz proposed in Ref. \cite{Nazir12} provides, in the Omhic case, a good approximation of both the GS and the ground energy of the SISBM.
\begin{figure*}[t!!] 
\begin{center}
{\includegraphics[width=0.3\textwidth]{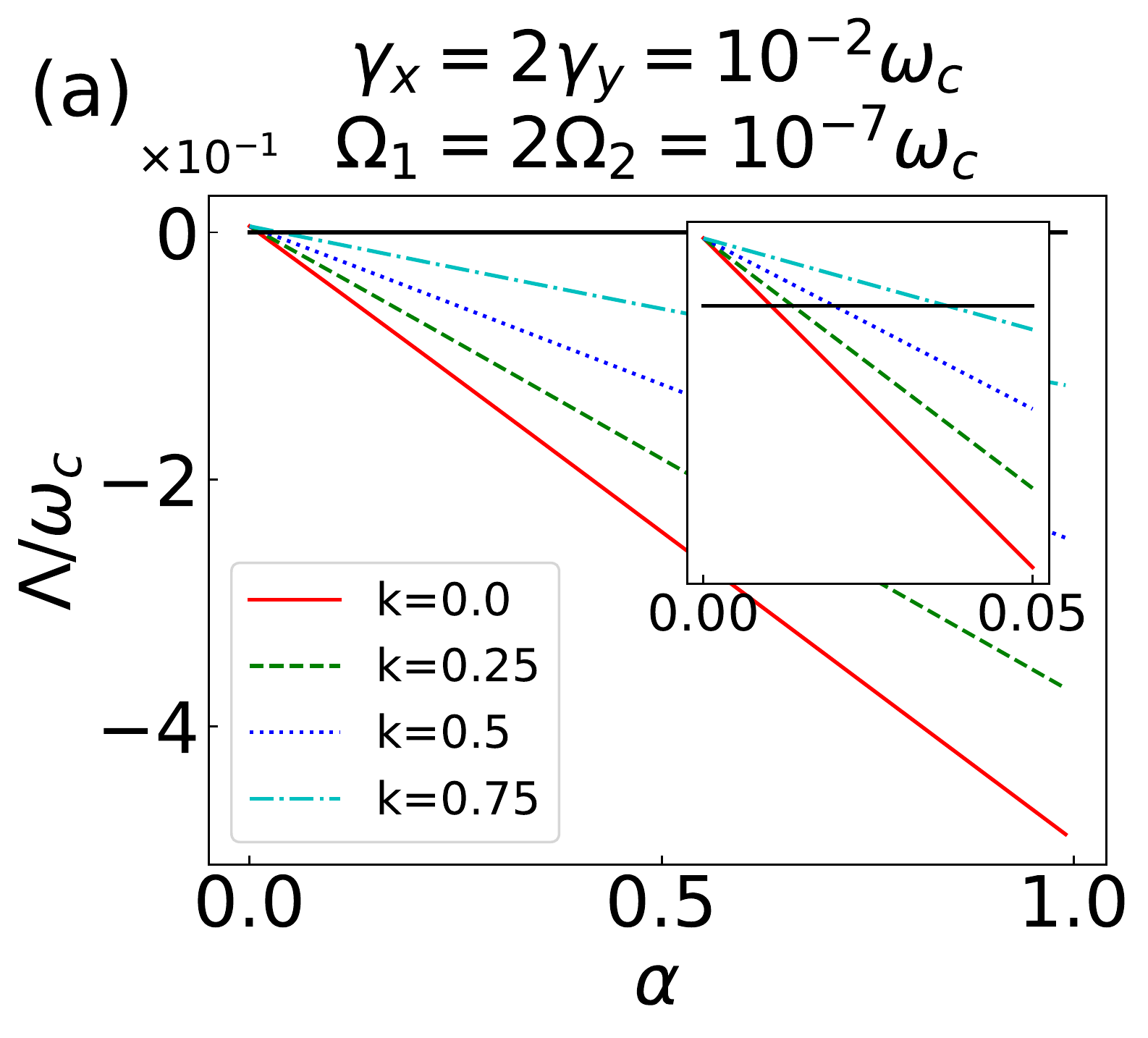}}
{\includegraphics[width=0.3\textwidth]{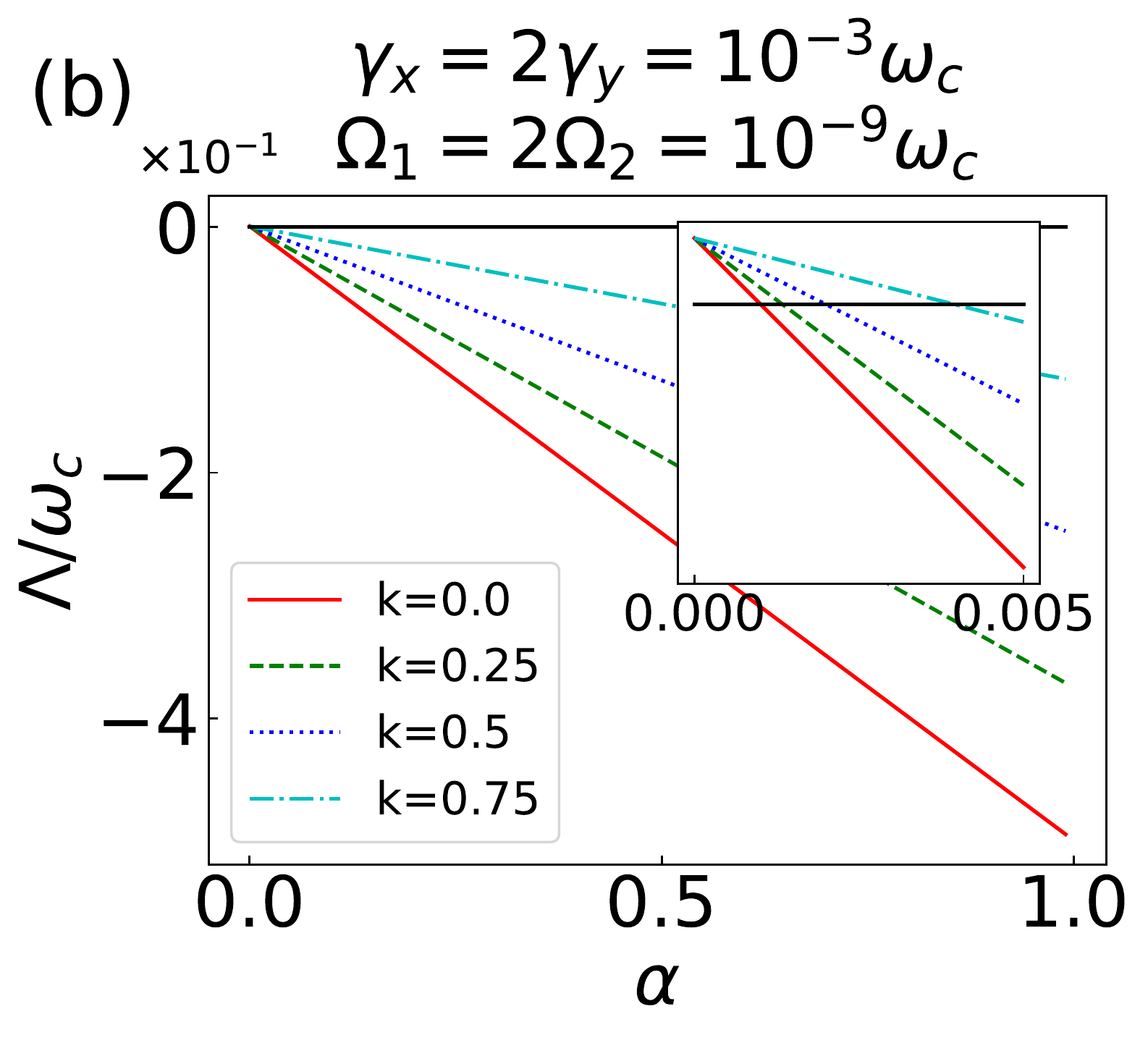}}
{\includegraphics[width=0.3\textwidth]{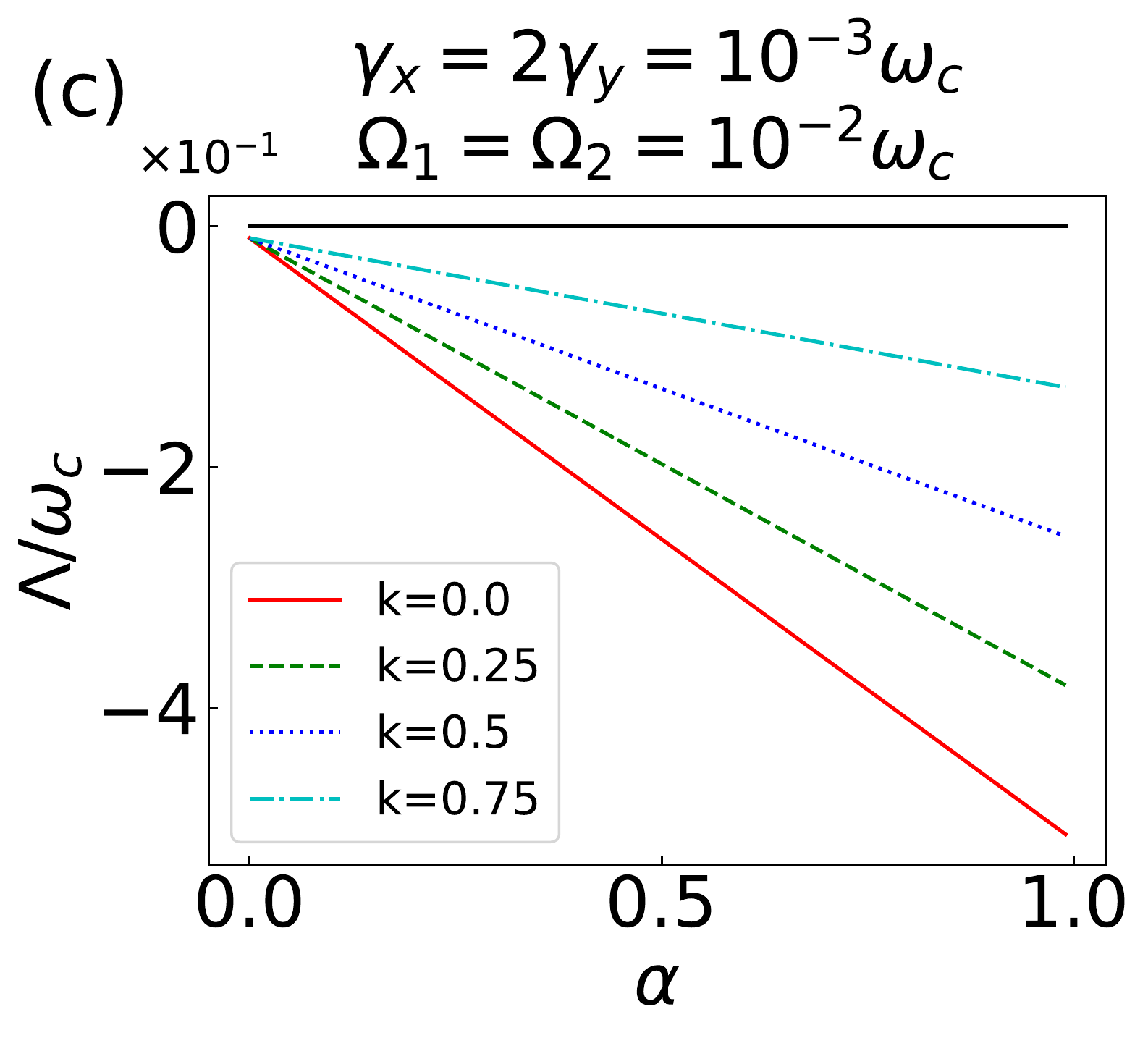}}
\captionsetup{justification=raggedright,format=plain,skip=4pt}%
\caption{Dependence of $\Lambda = \lambda_0^a - \lambda_0^b$ on $\alpha=\alpha_a=\alpha_b/k$. The black horizontal line represents $\Lambda=0$.
%for: (a) $\gamma_x=2\gamma_y=10^{-2}\omega_c$ and $\Omega_1=2\Omega_2=10^{-7}\omega_c$; (b) $\gamma_x=2\gamma_y=10^{-3}\omega_c$ and $\Omega_1=2\Omega_2=10^{-9}\omega_c$; (c) $\gamma_x=2\gamma_y=10^{-3}\omega_c$ and $\Omega_1=2\Omega_2=10^{-2}\omega_c$; with $k=0$ (solid red line), $k=0.25$ (dashed green line), $k=0.5$ (dotted blue line), $k=0.75$ (dot-dashed cyan line)
} \label{fig: GS}
\end{center}
\end{figure*}
Basing on these results, the GS of the TISBM, in the Omhic regime, results to have one of the following forms
\begin{subequations}
  \begin{align}
    \ket{GS}_{a} &= A_{a} \left[ \ket{++} \prod_k \ket{0_k^+} \right] + B_{a} \left[ \ket{--} \prod_k \ket{0_k^-} \right], \\
    \ket{GS}_{b} &= A_{b} \left[ \ket{+-} \prod_k \ket{0_k^+} \right] + B_{b} \left[ \ket{-+} \prod_k \ket{0_k^-} \right],
  \end{align}
\end{subequations}
where $\ket{0_k^\pm} = D(\beta_k^\pm) \ket{0}$ \cite{Nazir12}; $\ket{0}$ stands for the GS of the quantum oscillator bath, $D(\beta_k) = \exp\{ \beta_k(\hat{a}_k^\dagger - \hat{a}_k) \}$ ($\beta_k$ real) are the bosonic displacement operators, and with
\begin{subequations}
  \begin{align}
    A_{a/b} &= -{ (1+R_{a/b}) \Omega_{a/b} - \eta_{a/b} \over \mathrm{N}_{a/b}},
    \qquad
    B_{a/b} = {\gamma_{a/b}' \over \mathrm{N}_{a/b}}, \\
    %\mathrm{N}_\pm &= \sqrt{(R_\pm + \Omega_{a/b} - \eta_\pm)^2 + (\gamma_{b/a}')^2}, \\
    R_{a/b} &= {2 \alpha_{a/b} \omega_c \over \chi_{a/b} + \omega_c},
    \qquad
    \chi_{a/b} = \sqrt{(\gamma_{a/b}')^2 + \Omega_{a/b}^2}, \\
    \eta_{a/b} &= \sqrt{(\gamma_{a/b}')^2 + \Omega_{a/b}^2 (1+R_{a/b})^2}, \\
    \gamma_{a/b}' &= \gamma_{a/b} \left( {\chi_{a/b} \over \chi_{a/b} + \omega_c} \right)^{\alpha_{a/b}} \exp\{ \alpha_{a/b} \omega_c / (\chi_{a/b} + \omega_c) \},
  \end{align}
\end{subequations}
$\mathrm{N}_{a/b}$ being the normalization factors.
The self-consistent equation for $\gamma_{b/a}'$, in the scaling limit $\chi_{a/b}/\omega_c \rightarrow 0$, reads
\begin{subequations}
  \begin{align}
    \gamma_{a/b}' &= \left( {\gamma_{a/b} e^{\alpha_{a/b}} \over \omega_c^{\alpha_{a/b}}} \right)^{1/1-\alpha_{a/b}}, & \Omega_{a/b} \ll  T_K^{a/b}, \\
    \gamma_{a/b}' &= \gamma_{a/b} \left( {\Omega_{a/b} \over \omega_c} \right)^{\alpha_{a/b}}, & \Omega_{a/b} \gg T_K^{a/b}, 
  \end{align}
\end{subequations}
where $T_K^{a/b} = \gamma_{a/b}(\gamma_{a/b}/D)^{\alpha_{a/b}/1-\alpha_{a/b}}$ is the Kondo energy which scales the energies of the system, with $D$ being a cutoff introduced to regularize the integral for the ground-state energy \cite{LeHur08, Nazir12}.
We underline that, in our case, the scaling Kondo energy strictly depends on the spin-spin interaction strength.

The related energies can be cast as follows
\begin{equation}
    \lambda_{0}^{a/b} = {1 \over 2} \left( {\alpha_{a/b} \omega_c (\Omega_{a/b}^2 - \chi_{a/b} \omega_c) \over \chi_{a/b}(\chi_{a/b} + \omega_c)} - \eta_{a/b} \right). \label{GE}
\end{equation}
Therefore the spectrum of the TISBM is constituted by two sets of eigenvalues, each of which is the spectrum of the related effective SISBM.

Since the latter does not present level crossing, by studying the difference $\Lambda \equiv \lambda_0^a - \lambda_0^b$ the subspace where the GS of the TSBS is placed can be deduced.
In Fig. \ref{fig: GS}(a)
%($\gamma_x=2\gamma_y=10^{-2}\omega_c$ and $\Omega_1=2\Omega_2=10^{-7}\omega_c$)
the dependence of $\Lambda$ on $\alpha=\alpha_a=\alpha_b/k$ for different values of the parameter $k \equiv \alpha_b/\alpha_a$ is shown.
The $\alpha$-dependent straight lines emerge by considering the expression of $\Lambda$ in the limits $\Omega_{a/b} \rightarrow 0$ and $\gamma_{a/b} \ll \omega_c$, which becomes
\begin{equation} \label{Straight Lines}
    \Lambda \approx { (k-1)\omega_c \alpha + [\gamma_b'(\alpha) - \gamma_a'(\alpha)]\over 2}.
\end{equation}

The presence of a QPT at $\alpha_c \in [0.01,0.05]$ is clearly visible, as well as the dependence of the critical point ($\alpha_c$) on $k$.
The GS of the TSBS is then placed in the $b$ ($a$) space for $\alpha < \alpha_c$ ($\alpha > \alpha_c$).
The critical value $\alpha_c$, for which such a QPT occurs, is sensitive to the Hamiltonian parameters, as shown in Fig. \ref{fig: GS}(b),
%($\gamma_x=2\gamma_y=10^{-3}\omega_c$ and $\Omega_1=2\Omega_2=10^{-9}\omega_c$), 
where $\alpha_c \in [0.001,0.005]$.
This circumstance can be traced back to the different value taken on in the two cases by the intercept of the straight lines in Eq. \eqref{Straight Lines}, which is $(\gamma_b-\gamma_a)/2=\gamma_y$,
%\begin{equation}
%    {\gamma_b'(\alpha=0) - \gamma_a'(\alpha=0) \over 2} = \gamma_y, \qquad \Omega_{a/b} \ll \gamma_{a/b},
%\end{equation}
with the slope remaining unchanged.
In the isotropic case $\gamma_x=\gamma_y$ ($\gamma_a=0$), the QPT is still present since the intercept is $\gamma_b/2$.

In the QPT the spin-pair moves from the vanishing ($b)$ to the non-vanishing ($a)$ magnetization subspace.
In this way, the total magnetization plays the role of the order parameter.
It jumps from zero to a constant value, which can be derived on the basis of the result obtained for the SISBM, namely \cite{LeHur08,Nazir12}:
\begin{subequations} \label{Mag GS +}
  \begin{align}
    \average{\hat{\Sigma}^z} = & - C_z(\alpha) {\Omega_a \over T_K^{a}}, \qquad \Omega_a \ll T_K^a, \\
    C_z(\alpha) =& {4e^{\beta \over 2(1-\alpha)} \over \sqrt{\pi}} {\Gamma[1+1/(2-2\alpha)] \over \Gamma[1+\alpha/(2-2\alpha)]}, \\
    \beta =& \alpha \ln(\alpha) + (1-\alpha) \ln(1-\alpha).
  \end{align}
\end{subequations}
The transition can be then classified as a first-order QPT \cite{Vojta03}.

When $\Omega_1,\Omega_2 \gg \gamma_x,\gamma_y$ as in Fig. \ref{fig: GS}(d),
%($\gamma_x=2\gamma_y=10^{-3}\omega_c$ and $\Omega_1=2\Omega_2=10^{-2}\omega_c$),
no QPT occurs: the GS of the TSBS is $\ket{GS}_a$ and the net magnetization expression is equal to Eq. \eqref{Mag GS +}.
This is due to the fact that, in this case, $\Lambda$ can be approximated as in Eq. \eqref{Straight Lines}, but this time the intercept takes on the negative value $-\Omega_2$.
%$\Lambda$ can be approximated as
%\begin{equation} \label{Straight Lines new}
%    \Lambda \approx { (k-1)\omega_c \alpha \over 2} - \Omega_2, \qquad \Omega_{a/b} \gg \gamma_{a/b},
%\end{equation}
%and the intercept is negative.
For homogeneous magnetic fields, $\Omega_1=\Omega_2$ ($\Omega_b=0$), the QPT is still absent since the intercept reads $-\Omega_a/2$.

In the limit $\Omega_1,\Omega_2 \rightarrow 0$, $\alpha=1$ is another critical point.
For the SISBM in the Ohmic case, by mapping the model onto the anisotropic Kondo model with bosonization techniques, it has been proved that the K-T transition is present at $\alpha=1$ (in the scaling limit and for vanishing $z$-magnetic field) \cite{Leggett87}.
This critical value of $\alpha$ separates a localized phase at $\alpha>1$ (the spin is in $\ket{+}$ or $\ket{-}$), characterized by a renormalized vanishing tunnel splitting, from a delocalized phase at $\alpha<1$ with an effective non-vanishing tunnel energy \cite{Leggett87}.
Therefore, basing on our approach, we can claim that at $\alpha=1$ the TSBS undergoes a quantum phase transition of the K-T type, with a consequent localization of the two spins in the state $\ket{++}$ or $\ket{--}$ (the fictitious spin-1/2 $a$ localizes in $\ket{+}_a$ or $\ket{-}_a$).
In our case, the tunneling parameter which renormalizes to 0 for $\alpha > 1$, causing the localization phenomenon, consists in the spin-spin coupling.
We point out that, with respect to the previous works \cite{Dolgitzer21}, we obtain a different critical value of $\alpha$ for which a K-T transition occurs in the TISBM.
However, it must be reminded that our TISBM is different from those until now analysed: the absence of an external transverse field applied to the spin-pair causes such a remarkable difference.
The interesting aspects of the present model are both the possibility of rigorously deriving the existence of a K-T transition at $\alpha=1$ and the fact that such a transition relies on the presence of a non-vanishing transverse spin-spin interaction.

In conclusion, the present work, besides reporting nontrivial dynamical effects emerging in the TISBM, has shown the potentiality of the exact approach used.
The latter, exploited in other parameter-space regions, as well as in the sub-Omhic and super-Omhic regimes, can lead to a plethora of new results based on those obtained for the SISBM.

\bibliography{biblio_spin-boson.bib}

\begin{thebibliography}{50}%
\makeatletter
\providecommand \@ifxundefined [1]{%
 \@ifx{#1\undefined}
}%
\providecommand \@ifnum [1]{%
 \ifnum #1\expandafter \@firstoftwo
 \else \expandafter \@secondoftwo
 \fi
}%
\providecommand \@ifx [1]{%
 \ifx #1\expandafter \@firstoftwo
 \else \expandafter \@secondoftwo
 \fi
}%
\providecommand \natexlab [1]{#1}%
\providecommand \enquote  [1]{``#1''}%
\providecommand \bibnamefont  [1]{#1}%
\providecommand \bibfnamefont [1]{#1}%
\providecommand \citenamefont [1]{#1}%
\providecommand \href@noop [0]{\@secondoftwo}%
\providecommand \href [0]{\begingroup \@sanitize@url \@href}%
\providecommand \@href[1]{\@@startlink{#1}\@@href}%
\providecommand \@@href[1]{\endgroup#1\@@endlink}%
\providecommand \@sanitize@url [0]{\catcode `\\12\catcode `\$12\catcode
  `\&12\catcode `\#12\catcode `\^12\catcode `\_12\catcode `\%12\relax}%
\providecommand \@@startlink[1]{}%
\providecommand \@@endlink[0]{}%
\providecommand \url  [0]{\begingroup\@sanitize@url \@url }%
\providecommand \@url [1]{\endgroup\@href {#1}{\urlprefix }}%
\providecommand \urlprefix  [0]{URL }%
\providecommand \Eprint [0]{\href }%
\providecommand \doibase [0]{http://dx.doi.org/}%
\providecommand \selectlanguage [0]{\@gobble}%
\providecommand \bibinfo  [0]{\@secondoftwo}%
\providecommand \bibfield  [0]{\@secondoftwo}%
\providecommand \translation [1]{[#1]}%
\providecommand \BibitemOpen [0]{}%
\providecommand \bibitemStop [0]{}%
\providecommand \bibitemNoStop [0]{.\EOS\space}%
\providecommand \EOS [0]{\spacefactor3000\relax}%
\providecommand \BibitemShut  [1]{\csname bibitem#1\endcsname}%
\let\auto@bib@innerbib\@empty
%</preamble>
\bibitem [{\citenamefont {Breuer}\ and\ \citenamefont
  {Petruccione}(2002)}]{BP02}%
  \BibitemOpen
  \bibfield  {author} {\bibinfo {author} {\bibfnamefont {H.-P.}\ \bibnamefont
  {Breuer}}\ and\ \bibinfo {author} {\bibfnamefont {F.}~\bibnamefont
  {Petruccione}},\ }\href@noop {} {\emph {\bibinfo {title} {The theory of open
  quantum systems}}}\ (\bibinfo  {publisher} {Oxford University Press on
  Demand},\ \bibinfo {year} {2002})\BibitemShut {NoStop}%
\bibitem [{\citenamefont {Nielsen}\ and\ \citenamefont {Chuang}(2010)}]{NC10}%
  \BibitemOpen
  \bibfield  {author} {\bibinfo {author} {\bibfnamefont {M.~A.}\ \bibnamefont
  {Nielsen}}\ and\ \bibinfo {author} {\bibfnamefont {I.~L.}\ \bibnamefont
  {Chuang}},\ }\href {\doibase 10.1017/CBO9780511976667} {\emph {\bibinfo
  {title} {Quantum Computation and Quantum Information}}}\ (\bibinfo
  {publisher} {Cambridge University Press},\ \bibinfo {year}
  {2010})\BibitemShut {NoStop}%
\bibitem [{\citenamefont {Leggett}\ \emph {et~al.}(1987)\citenamefont
  {Leggett}, \citenamefont {Chakravarty}, \citenamefont {Dorsey}, \citenamefont
  {Fisher}, \citenamefont {Garg},\ and\ \citenamefont {Zwerger}}]{Leggett87}%
  \BibitemOpen
  \bibfield  {author} {\bibinfo {author} {\bibfnamefont {A.~J.}\ \bibnamefont
  {Leggett}}, \bibinfo {author} {\bibfnamefont {S.}~\bibnamefont
  {Chakravarty}}, \bibinfo {author} {\bibfnamefont {A.~T.}\ \bibnamefont
  {Dorsey}}, \bibinfo {author} {\bibfnamefont {M.~P.~A.}\ \bibnamefont
  {Fisher}}, \bibinfo {author} {\bibfnamefont {A.}~\bibnamefont {Garg}}, \ and\
  \bibinfo {author} {\bibfnamefont {W.}~\bibnamefont {Zwerger}},\ }\href
  {\doibase 10.1103/RevModPhys.59.1} {\bibfield  {journal} {\bibinfo  {journal}
  {Rev. Mod. Phys.}\ }\textbf {\bibinfo {volume} {59}},\ \bibinfo {pages} {1}
  (\bibinfo {year} {1987})}\BibitemShut {NoStop}%
\bibitem [{\citenamefont {Le~Hur}\ \emph {et~al.}(2007)\citenamefont {Le~Hur},
  \citenamefont {Doucet-Beaupr\'e},\ and\ \citenamefont
  {Hofstetter}}]{LeHur07}%
  \BibitemOpen
  \bibfield  {author} {\bibinfo {author} {\bibfnamefont {K.}~\bibnamefont
  {Le~Hur}}, \bibinfo {author} {\bibfnamefont {P.}~\bibnamefont
  {Doucet-Beaupr\'e}}, \ and\ \bibinfo {author} {\bibfnamefont
  {W.}~\bibnamefont {Hofstetter}},\ }\href {\doibase
  10.1103/PhysRevLett.99.126801} {\bibfield  {journal} {\bibinfo  {journal}
  {Phys. Rev. Lett.}\ }\textbf {\bibinfo {volume} {99}},\ \bibinfo {pages}
  {126801} (\bibinfo {year} {2007})}\BibitemShut {NoStop}%
\bibitem [{\citenamefont {Vojta}\ \emph {et~al.}(2005)\citenamefont {Vojta},
  \citenamefont {Tong},\ and\ \citenamefont {Bulla}}]{Vojta05}%
  \BibitemOpen
  \bibfield  {author} {\bibinfo {author} {\bibfnamefont {M.}~\bibnamefont
  {Vojta}}, \bibinfo {author} {\bibfnamefont {N.-H.}\ \bibnamefont {Tong}}, \
  and\ \bibinfo {author} {\bibfnamefont {R.}~\bibnamefont {Bulla}},\ }\href
  {\doibase 10.1103/PhysRevLett.94.070604} {\bibfield  {journal} {\bibinfo
  {journal} {Phys. Rev. Lett.}\ }\textbf {\bibinfo {volume} {94}},\ \bibinfo
  {pages} {070604} (\bibinfo {year} {2005})}\BibitemShut {NoStop}%
\bibitem [{\citenamefont {Bulla}\ \emph {et~al.}(2003)\citenamefont {Bulla},
  \citenamefont {Tong},\ and\ \citenamefont {Vojta}}]{Bulla03}%
  \BibitemOpen
  \bibfield  {author} {\bibinfo {author} {\bibfnamefont {R.}~\bibnamefont
  {Bulla}}, \bibinfo {author} {\bibfnamefont {N.-H.}\ \bibnamefont {Tong}}, \
  and\ \bibinfo {author} {\bibfnamefont {M.}~\bibnamefont {Vojta}},\ }\href
  {\doibase 10.1103/PhysRevLett.91.170601} {\bibfield  {journal} {\bibinfo
  {journal} {Phys. Rev. Lett.}\ }\textbf {\bibinfo {volume} {91}},\ \bibinfo
  {pages} {170601} (\bibinfo {year} {2003})}\BibitemShut {NoStop}%
\bibitem [{\citenamefont {Hur}(2008)}]{LeHur08}%
  \BibitemOpen
  \bibfield  {author} {\bibinfo {author} {\bibfnamefont {K.~L.}\ \bibnamefont
  {Hur}},\ }\href {\doibase https://doi.org/10.1016/j.aop.2007.12.003}
  {\bibfield  {journal} {\bibinfo  {journal} {Ann. Phys. (NY)}\ }\textbf
  {\bibinfo {volume} {323}},\ \bibinfo {pages} {2208} (\bibinfo {year}
  {2008})}\BibitemShut {NoStop}%
\bibitem [{\citenamefont {Nazir}\ \emph {et~al.}(2012)\citenamefont {Nazir},
  \citenamefont {McCutcheon},\ and\ \citenamefont {Chin}}]{Nazir12}%
  \BibitemOpen
  \bibfield  {author} {\bibinfo {author} {\bibfnamefont {A.}~\bibnamefont
  {Nazir}}, \bibinfo {author} {\bibfnamefont {D.~P.~S.}\ \bibnamefont
  {McCutcheon}}, \ and\ \bibinfo {author} {\bibfnamefont {A.~W.}\ \bibnamefont
  {Chin}},\ }\href {\doibase 10.1103/PhysRevB.85.224301} {\bibfield  {journal}
  {\bibinfo  {journal} {Phys. Rev. B}\ }\textbf {\bibinfo {volume} {85}},\
  \bibinfo {pages} {224301} (\bibinfo {year} {2012})}\BibitemShut {NoStop}%
\bibitem [{\citenamefont {Vojta}(2003)}]{Vojta03}%
  \BibitemOpen
  \bibfield  {author} {\bibinfo {author} {\bibfnamefont {M.}~\bibnamefont
  {Vojta}},\ }\href {\doibase 10.1088/0034-4885/66/12/r01} {\bibfield
  {journal} {\bibinfo  {journal} {Rep. Prog. Phys.}\ }\textbf {\bibinfo
  {volume} {66}},\ \bibinfo {pages} {2069} (\bibinfo {year}
  {2003})}\BibitemShut {NoStop}%
\bibitem [{\citenamefont {Rossini}\ and\ \citenamefont
  {Vicari}(2021)}]{Rossini21}%
  \BibitemOpen
  \bibfield  {author} {\bibinfo {author} {\bibfnamefont {D.}~\bibnamefont
  {Rossini}}\ and\ \bibinfo {author} {\bibfnamefont {E.}~\bibnamefont
  {Vicari}},\ }\href {\doibase https://doi.org/10.1016/j.physrep.2021.08.003}
  {\bibfield  {journal} {\bibinfo  {journal} {Phys. Rep.}\ }\textbf {\bibinfo
  {volume} {936}},\ \bibinfo {pages} {1} (\bibinfo {year} {2021})},\ \bibinfo
  {note} {coherent and dissipative dynamics at quantum phase
  transitions}\BibitemShut {NoStop}%
\bibitem [{\citenamefont {Carollo}\ \emph {et~al.}(2020)\citenamefont
  {Carollo}, \citenamefont {Valenti},\ and\ \citenamefont
  {Spagnolo}}]{Carollo20}%
  \BibitemOpen
  \bibfield  {author} {\bibinfo {author} {\bibfnamefont {A.}~\bibnamefont
  {Carollo}}, \bibinfo {author} {\bibfnamefont {D.}~\bibnamefont {Valenti}}, \
  and\ \bibinfo {author} {\bibfnamefont {B.}~\bibnamefont {Spagnolo}},\ }\href
  {\doibase https://doi.org/10.1016/j.physrep.2019.11.002} {\bibfield
  {journal} {\bibinfo  {journal} {Phys. Rep.}\ }\textbf {\bibinfo {volume}
  {838}},\ \bibinfo {pages} {1} (\bibinfo {year} {2020})}\BibitemShut {NoStop}%
\bibitem [{\citenamefont {Kundu}\ and\ \citenamefont {Makri}(2021)}]{Kundu21}%
  \BibitemOpen
  \bibfield  {author} {\bibinfo {author} {\bibfnamefont {S.}~\bibnamefont
  {Kundu}}\ and\ \bibinfo {author} {\bibfnamefont {N.}~\bibnamefont {Makri}},\
  }\href {\doibase 10.1021/acs.jpcb.1c03861} {\bibfield  {journal} {\bibinfo
  {journal} {J. Phys. Chem. B}\ }\textbf {\bibinfo {volume} {125}},\ \bibinfo
  {pages} {8137} (\bibinfo {year} {2021})}\BibitemShut {NoStop}%
\bibitem [{\citenamefont {Dunnett}\ and\ \citenamefont
  {Chin}(2021)}]{Dunnett21}%
  \BibitemOpen
  \bibfield  {author} {\bibinfo {author} {\bibfnamefont {A.~J.}\ \bibnamefont
  {Dunnett}}\ and\ \bibinfo {author} {\bibfnamefont {A.~W.}\ \bibnamefont
  {Chin}},\ }\href {\doibase 10.3390/e23010077} {\bibfield  {journal} {\bibinfo
   {journal} {Entropy}\ }\textbf {\bibinfo {volume} {23}} (\bibinfo {year}
  {2021}),\ 10.3390/e23010077}\BibitemShut {NoStop}%
\bibitem [{\citenamefont {Lemmer}\ \emph {et~al.}(2018)\citenamefont {Lemmer},
  \citenamefont {Cormick}, \citenamefont {Tamascelli}, \citenamefont {Schaetz},
  \citenamefont {Huelga},\ and\ \citenamefont {Plenio}}]{Lemmer18}%
  \BibitemOpen
  \bibfield  {author} {\bibinfo {author} {\bibfnamefont {A.}~\bibnamefont
  {Lemmer}}, \bibinfo {author} {\bibfnamefont {C.}~\bibnamefont {Cormick}},
  \bibinfo {author} {\bibfnamefont {D.}~\bibnamefont {Tamascelli}}, \bibinfo
  {author} {\bibfnamefont {T.}~\bibnamefont {Schaetz}}, \bibinfo {author}
  {\bibfnamefont {S.~F.}\ \bibnamefont {Huelga}}, \ and\ \bibinfo {author}
  {\bibfnamefont {M.~B.}\ \bibnamefont {Plenio}},\ }\href {\doibase
  10.1088/1367-2630/aac87d} {\bibfield  {journal} {\bibinfo  {journal} {New J.
  Phys.}\ }\textbf {\bibinfo {volume} {20}},\ \bibinfo {pages} {073002}
  (\bibinfo {year} {2018})}\BibitemShut {NoStop}%
\bibitem [{\citenamefont {Lerma-Hern\'andez}\ \emph {et~al.}(2019)\citenamefont
  {Lerma-Hern\'andez}, \citenamefont {Villase\~nor}, \citenamefont
  {Bastarrachea-Magnani}, \citenamefont {Torres-Herrera}, \citenamefont
  {Santos},\ and\ \citenamefont {Hirsch}}]{Lerma19}%
  \BibitemOpen
  \bibfield  {author} {\bibinfo {author} {\bibfnamefont {S.}~\bibnamefont
  {Lerma-Hern\'andez}}, \bibinfo {author} {\bibfnamefont {D.}~\bibnamefont
  {Villase\~nor}}, \bibinfo {author} {\bibfnamefont {M.~A.}\ \bibnamefont
  {Bastarrachea-Magnani}}, \bibinfo {author} {\bibfnamefont {E.~J.}\
  \bibnamefont {Torres-Herrera}}, \bibinfo {author} {\bibfnamefont {L.~F.}\
  \bibnamefont {Santos}}, \ and\ \bibinfo {author} {\bibfnamefont {J.~G.}\
  \bibnamefont {Hirsch}},\ }\href {\doibase 10.1103/PhysRevE.100.012218}
  {\bibfield  {journal} {\bibinfo  {journal} {Phys. Rev. E}\ }\textbf {\bibinfo
  {volume} {100}},\ \bibinfo {pages} {012218} (\bibinfo {year}
  {2019})}\BibitemShut {NoStop}%
\bibitem [{\citenamefont {Lepp\"akangas}\ \emph {et~al.}(2018)\citenamefont
  {Lepp\"akangas}, \citenamefont {Braum\"uller}, \citenamefont {Hauck},
  \citenamefont {Reiner}, \citenamefont {Schwenk}, \citenamefont {Zanker},
  \citenamefont {Fritz}, \citenamefont {Ustinov}, \citenamefont {Weides},\ and\
  \citenamefont {Marthaler}}]{Leppakangas18}%
  \BibitemOpen
  \bibfield  {author} {\bibinfo {author} {\bibfnamefont {J.}~\bibnamefont
  {Lepp\"akangas}}, \bibinfo {author} {\bibfnamefont {J.}~\bibnamefont
  {Braum\"uller}}, \bibinfo {author} {\bibfnamefont {M.}~\bibnamefont {Hauck}},
  \bibinfo {author} {\bibfnamefont {J.-M.}\ \bibnamefont {Reiner}}, \bibinfo
  {author} {\bibfnamefont {I.}~\bibnamefont {Schwenk}}, \bibinfo {author}
  {\bibfnamefont {S.}~\bibnamefont {Zanker}}, \bibinfo {author} {\bibfnamefont
  {L.}~\bibnamefont {Fritz}}, \bibinfo {author} {\bibfnamefont {A.~V.}\
  \bibnamefont {Ustinov}}, \bibinfo {author} {\bibfnamefont {M.}~\bibnamefont
  {Weides}}, \ and\ \bibinfo {author} {\bibfnamefont {M.}~\bibnamefont
  {Marthaler}},\ }\href {\doibase 10.1103/PhysRevA.97.052321} {\bibfield
  {journal} {\bibinfo  {journal} {Phys. Rev. A}\ }\textbf {\bibinfo {volume}
  {97}},\ \bibinfo {pages} {052321} (\bibinfo {year} {2018})}\BibitemShut
  {NoStop}%
\bibitem [{\citenamefont {Puebla}\ \emph {et~al.}(2019)\citenamefont {Puebla},
  \citenamefont {Casanova}, \citenamefont {Houhou}, \citenamefont {Solano},\
  and\ \citenamefont {Paternostro}}]{Puebla19}%
  \BibitemOpen
  \bibfield  {author} {\bibinfo {author} {\bibfnamefont {R.}~\bibnamefont
  {Puebla}}, \bibinfo {author} {\bibfnamefont {J.}~\bibnamefont {Casanova}},
  \bibinfo {author} {\bibfnamefont {O.}~\bibnamefont {Houhou}}, \bibinfo
  {author} {\bibfnamefont {E.}~\bibnamefont {Solano}}, \ and\ \bibinfo {author}
  {\bibfnamefont {M.}~\bibnamefont {Paternostro}},\ }\href {\doibase
  10.1103/PhysRevA.99.032303} {\bibfield  {journal} {\bibinfo  {journal} {Phys.
  Rev. A}\ }\textbf {\bibinfo {volume} {99}},\ \bibinfo {pages} {032303}
  (\bibinfo {year} {2019})}\BibitemShut {NoStop}%
\bibitem [{\citenamefont {Wenderoth}\ \emph {et~al.}(2021)\citenamefont
  {Wenderoth}, \citenamefont {Breuer},\ and\ \citenamefont
  {Thoss}}]{Wenderoth21}%
  \BibitemOpen
  \bibfield  {author} {\bibinfo {author} {\bibfnamefont {S.}~\bibnamefont
  {Wenderoth}}, \bibinfo {author} {\bibfnamefont {H.-P.}\ \bibnamefont
  {Breuer}}, \ and\ \bibinfo {author} {\bibfnamefont {M.}~\bibnamefont
  {Thoss}},\ }\href {\doibase 10.1103/PhysRevA.104.012213} {\bibfield
  {journal} {\bibinfo  {journal} {Phys. Rev. A}\ }\textbf {\bibinfo {volume}
  {104}},\ \bibinfo {pages} {012213} (\bibinfo {year} {2021})}\BibitemShut
  {NoStop}%
\bibitem [{\citenamefont {Magazz\`u}\ \emph {et~al.}(2018)\citenamefont
  {Magazz\`u}, \citenamefont {Denisov},\ and\ \citenamefont
  {H\"anggi}}]{Magazzu18}%
  \BibitemOpen
  \bibfield  {author} {\bibinfo {author} {\bibfnamefont {L.}~\bibnamefont
  {Magazz\`u}}, \bibinfo {author} {\bibfnamefont {S.}~\bibnamefont {Denisov}},
  \ and\ \bibinfo {author} {\bibfnamefont {P.}~\bibnamefont {H\"anggi}},\
  }\href {\doibase 10.1103/PhysRevE.98.022111} {\bibfield  {journal} {\bibinfo
  {journal} {Phys. Rev. E}\ }\textbf {\bibinfo {volume} {98}},\ \bibinfo
  {pages} {022111} (\bibinfo {year} {2018})}\BibitemShut {NoStop}%
\bibitem [{\citenamefont {De~Filippis}\ \emph {et~al.}(2020)\citenamefont
  {De~Filippis}, \citenamefont {de~Candia}, \citenamefont {Cangemi},
  \citenamefont {Sassetti}, \citenamefont {Fazio},\ and\ \citenamefont
  {Cataudella}}]{DeFilippis20}%
  \BibitemOpen
  \bibfield  {author} {\bibinfo {author} {\bibfnamefont {G.}~\bibnamefont
  {De~Filippis}}, \bibinfo {author} {\bibfnamefont {A.}~\bibnamefont
  {de~Candia}}, \bibinfo {author} {\bibfnamefont {L.~M.}\ \bibnamefont
  {Cangemi}}, \bibinfo {author} {\bibfnamefont {M.}~\bibnamefont {Sassetti}},
  \bibinfo {author} {\bibfnamefont {R.}~\bibnamefont {Fazio}}, \ and\ \bibinfo
  {author} {\bibfnamefont {V.}~\bibnamefont {Cataudella}},\ }\href {\doibase
  10.1103/PhysRevB.101.180408} {\bibfield  {journal} {\bibinfo  {journal}
  {Phys. Rev. B}\ }\textbf {\bibinfo {volume} {101}},\ \bibinfo {pages}
  {180408} (\bibinfo {year} {2020})}\BibitemShut {NoStop}%
\bibitem [{\citenamefont {Wang}\ \emph {et~al.}(2019)\citenamefont {Wang},
  \citenamefont {He}, \citenamefont {Duan},\ and\ \citenamefont
  {Chen}}]{Wang19}%
  \BibitemOpen
  \bibfield  {author} {\bibinfo {author} {\bibfnamefont {Y.-Z.}\ \bibnamefont
  {Wang}}, \bibinfo {author} {\bibfnamefont {S.}~\bibnamefont {He}}, \bibinfo
  {author} {\bibfnamefont {L.}~\bibnamefont {Duan}}, \ and\ \bibinfo {author}
  {\bibfnamefont {Q.-H.}\ \bibnamefont {Chen}},\ }\href {\doibase
  10.1103/PhysRevB.100.115106} {\bibfield  {journal} {\bibinfo  {journal}
  {Phys. Rev. B}\ }\textbf {\bibinfo {volume} {100}},\ \bibinfo {pages}
  {115106} (\bibinfo {year} {2019})}\BibitemShut {NoStop}%
\bibitem [{\citenamefont {Wang}\ \emph {et~al.}(2020)\citenamefont {Wang},
  \citenamefont {He}, \citenamefont {Duan},\ and\ \citenamefont
  {Chen}}]{Wang20}%
  \BibitemOpen
  \bibfield  {author} {\bibinfo {author} {\bibfnamefont {Y.-Z.}\ \bibnamefont
  {Wang}}, \bibinfo {author} {\bibfnamefont {S.}~\bibnamefont {He}}, \bibinfo
  {author} {\bibfnamefont {L.}~\bibnamefont {Duan}}, \ and\ \bibinfo {author}
  {\bibfnamefont {Q.-H.}\ \bibnamefont {Chen}},\ }\href {\doibase
  10.1103/PhysRevB.101.155147} {\bibfield  {journal} {\bibinfo  {journal}
  {Phys. Rev. B}\ }\textbf {\bibinfo {volume} {101}},\ \bibinfo {pages}
  {155147} (\bibinfo {year} {2020})}\BibitemShut {NoStop}%
\bibitem [{\citenamefont {Shen}\ \emph {et~al.}(2021)\citenamefont {Shen},
  \citenamefont {Yang}, \citenamefont {Zhong}, \citenamefont {Yang},\ and\
  \citenamefont {Zheng}}]{Shen21}%
  \BibitemOpen
  \bibfield  {author} {\bibinfo {author} {\bibfnamefont {L.-T.}\ \bibnamefont
  {Shen}}, \bibinfo {author} {\bibfnamefont {J.-W.}\ \bibnamefont {Yang}},
  \bibinfo {author} {\bibfnamefont {Z.-R.}\ \bibnamefont {Zhong}}, \bibinfo
  {author} {\bibfnamefont {Z.-B.}\ \bibnamefont {Yang}}, \ and\ \bibinfo
  {author} {\bibfnamefont {S.-B.}\ \bibnamefont {Zheng}},\ }\href {\doibase
  10.1103/PhysRevA.104.063703} {\bibfield  {journal} {\bibinfo  {journal}
  {Phys. Rev. A}\ }\textbf {\bibinfo {volume} {104}},\ \bibinfo {pages}
  {063703} (\bibinfo {year} {2021})}\BibitemShut {NoStop}%
\bibitem [{\citenamefont {Aurell}\ \emph {et~al.}(2020)\citenamefont {Aurell},
  \citenamefont {Donvil},\ and\ \citenamefont {Mallick}}]{Aurell20}%
  \BibitemOpen
  \bibfield  {author} {\bibinfo {author} {\bibfnamefont {E.}~\bibnamefont
  {Aurell}}, \bibinfo {author} {\bibfnamefont {B.}~\bibnamefont {Donvil}}, \
  and\ \bibinfo {author} {\bibfnamefont {K.}~\bibnamefont {Mallick}},\ }\href
  {\doibase 10.1103/PhysRevE.101.052116} {\bibfield  {journal} {\bibinfo
  {journal} {Phys. Rev. E}\ }\textbf {\bibinfo {volume} {101}},\ \bibinfo
  {pages} {052116} (\bibinfo {year} {2020})}\BibitemShut {NoStop}%
\bibitem [{\citenamefont {Miessen}\ \emph {et~al.}(2021)\citenamefont
  {Miessen}, \citenamefont {Ollitrault},\ and\ \citenamefont
  {Tavernelli}}]{Miessen21}%
  \BibitemOpen
  \bibfield  {author} {\bibinfo {author} {\bibfnamefont {A.}~\bibnamefont
  {Miessen}}, \bibinfo {author} {\bibfnamefont {P.~J.}\ \bibnamefont
  {Ollitrault}}, \ and\ \bibinfo {author} {\bibfnamefont {I.}~\bibnamefont
  {Tavernelli}},\ }\href {\doibase 10.1103/PhysRevResearch.3.043212} {\bibfield
   {journal} {\bibinfo  {journal} {Phys. Rev. Research}\ }\textbf {\bibinfo
  {volume} {3}},\ \bibinfo {pages} {043212} (\bibinfo {year}
  {2021})}\BibitemShut {NoStop}%
\bibitem [{\citenamefont {Villase{\~{n}}or}\ \emph {et~al.}(2020)\citenamefont
  {Villase{\~{n}}or}, \citenamefont {Pilatowsky-Cameo}, \citenamefont
  {Bastarrachea-Magnani}, \citenamefont {Lerma-Hern{\'{a}}ndez}, \citenamefont
  {Santos},\ and\ \citenamefont {Hirsch}}]{Villase20}%
  \BibitemOpen
  \bibfield  {author} {\bibinfo {author} {\bibfnamefont {D.}~\bibnamefont
  {Villase{\~{n}}or}}, \bibinfo {author} {\bibfnamefont {S.}~\bibnamefont
  {Pilatowsky-Cameo}}, \bibinfo {author} {\bibfnamefont {M.~A.}\ \bibnamefont
  {Bastarrachea-Magnani}}, \bibinfo {author} {\bibfnamefont {S.}~\bibnamefont
  {Lerma-Hern{\'{a}}ndez}}, \bibinfo {author} {\bibfnamefont {L.~F.}\
  \bibnamefont {Santos}}, \ and\ \bibinfo {author} {\bibfnamefont {J.~G.}\
  \bibnamefont {Hirsch}},\ }\href {\doibase 10.1088/1367-2630/ab8ef8}
  {\bibfield  {journal} {\bibinfo  {journal} {New J. Phys.}\ }\textbf {\bibinfo
  {volume} {22}},\ \bibinfo {pages} {063036} (\bibinfo {year}
  {2020})}\BibitemShut {NoStop}%
\bibitem [{\citenamefont {Pino}\ and\ \citenamefont
  {Garc{\'{\i}}a-Ripoll}(2018)}]{Pino18}%
  \BibitemOpen
  \bibfield  {author} {\bibinfo {author} {\bibfnamefont {M.}~\bibnamefont
  {Pino}}\ and\ \bibinfo {author} {\bibfnamefont {J.~J.}\ \bibnamefont
  {Garc{\'{\i}}a-Ripoll}},\ }\href {\doibase 10.1088/1367-2630/aaeeea}
  {\bibfield  {journal} {\bibinfo  {journal} {New J. Phys.}\ }\textbf {\bibinfo
  {volume} {20}},\ \bibinfo {pages} {113027} (\bibinfo {year}
  {2018})}\BibitemShut {NoStop}%
\bibitem [{\citenamefont {Magazzù}\ \emph {et~al.}(2018)\citenamefont
  {Magazzù}, \citenamefont {Forn-Díaz}, \citenamefont {Belyansky},\ and\
  \citenamefont {\textit{et al.}}}]{Magazzu18nat}%
  \BibitemOpen
  \bibfield  {author} {\bibinfo {author} {\bibfnamefont {L.}~\bibnamefont
  {Magazzù}}, \bibinfo {author} {\bibfnamefont {P.}~\bibnamefont
  {Forn-Díaz}}, \bibinfo {author} {\bibfnamefont {R.}~\bibnamefont
  {Belyansky}}, \ and\ \bibinfo {author} {\bibnamefont {\textit{et al.}}},\
  }\href {\doibase 10.1038/s41467-018-03626-w} {\bibfield  {journal} {\bibinfo
  {journal} {Nat. Commun.}\ }\textbf {\bibinfo {volume} {9}} (\bibinfo {year}
  {2018}),\ 10.1038/s41467-018-03626-w}\BibitemShut {NoStop}%
\bibitem [{\citenamefont {Lambert}\ \emph {et~al.}(2019)\citenamefont
  {Lambert}, \citenamefont {Ahmed}, \citenamefont {Cirio},\ and\ \citenamefont
  {\textit{et al.}}}]{Lambert19}%
  \BibitemOpen
  \bibfield  {author} {\bibinfo {author} {\bibfnamefont {N.}~\bibnamefont
  {Lambert}}, \bibinfo {author} {\bibfnamefont {S.}~\bibnamefont {Ahmed}},
  \bibinfo {author} {\bibfnamefont {M.}~\bibnamefont {Cirio}}, \ and\ \bibinfo
  {author} {\bibnamefont {\textit{et al.}}},\ }\href {\doibase
  10.1038/s41467-019-11656-1} {\bibfield  {journal} {\bibinfo  {journal} {Nat.
  Commun.}\ }\textbf {\bibinfo {volume} {10}} (\bibinfo {year} {2019}),\
  10.1038/s41467-019-11656-1}\BibitemShut {NoStop}%
\bibitem [{\citenamefont {Casanova}\ \emph {et~al.}(2018)\citenamefont
  {Casanova}, \citenamefont {Puebla}, \citenamefont {Moya-Cessa},\ and\
  \citenamefont {\textit{et al.}}}]{Casanova18}%
  \BibitemOpen
  \bibfield  {author} {\bibinfo {author} {\bibfnamefont {J.}~\bibnamefont
  {Casanova}}, \bibinfo {author} {\bibfnamefont {R.}~\bibnamefont {Puebla}},
  \bibinfo {author} {\bibfnamefont {H.}~\bibnamefont {Moya-Cessa}}, \ and\
  \bibinfo {author} {\bibnamefont {\textit{et al.}}},\ }\href {\doibase
  10.1038/s41534-018-0096-9} {\bibfield  {journal} {\bibinfo  {journal} {npj
  Quantum. Inf.}\ }\textbf {\bibinfo {volume} {4}} (\bibinfo {year} {2018}),\
  10.1038/s41534-018-0096-9}\BibitemShut {NoStop}%
\bibitem [{\citenamefont {Dolgitzer}\ \emph {et~al.}(2021)\citenamefont
  {Dolgitzer}, \citenamefont {Zeng},\ and\ \citenamefont {Chen}}]{Dolgitzer21}%
  \BibitemOpen
  \bibfield  {author} {\bibinfo {author} {\bibfnamefont {D.}~\bibnamefont
  {Dolgitzer}}, \bibinfo {author} {\bibfnamefont {D.}~\bibnamefont {Zeng}}, \
  and\ \bibinfo {author} {\bibfnamefont {Y.}~\bibnamefont {Chen}},\ }\href
  {\doibase 10.1364/OE.434183} {\bibfield  {journal} {\bibinfo  {journal} {Opt.
  Express}\ }\textbf {\bibinfo {volume} {29}},\ \bibinfo {pages} {23988}
  (\bibinfo {year} {2021})}\BibitemShut {NoStop}%
\bibitem [{\citenamefont {Wang}\ \emph {et~al.}(2021)\citenamefont {Wang},
  \citenamefont {He}, \citenamefont {Duan},\ and\ \citenamefont
  {Chen}}]{Wang21}%
  \BibitemOpen
  \bibfield  {author} {\bibinfo {author} {\bibfnamefont {Y.-Z.}\ \bibnamefont
  {Wang}}, \bibinfo {author} {\bibfnamefont {S.}~\bibnamefont {He}}, \bibinfo
  {author} {\bibfnamefont {L.}~\bibnamefont {Duan}}, \ and\ \bibinfo {author}
  {\bibfnamefont {Q.-H.}\ \bibnamefont {Chen}},\ }\href {\doibase
  10.1103/PhysRevB.103.205106} {\bibfield  {journal} {\bibinfo  {journal}
  {Phys. Rev. B}\ }\textbf {\bibinfo {volume} {103}},\ \bibinfo {pages}
  {205106} (\bibinfo {year} {2021})}\BibitemShut {NoStop}%
\bibitem [{\citenamefont {Zhou}\ \emph {et~al.}(2018)\citenamefont {Zhou},
  \citenamefont {Zhang}, \citenamefont {Lü},\ and\ \citenamefont
  {Zhao}}]{Zhou18}%
  \BibitemOpen
  \bibfield  {author} {\bibinfo {author} {\bibfnamefont {N.}~\bibnamefont
  {Zhou}}, \bibinfo {author} {\bibfnamefont {Y.}~\bibnamefont {Zhang}},
  \bibinfo {author} {\bibfnamefont {Z.}~\bibnamefont {Lü}}, \ and\ \bibinfo
  {author} {\bibfnamefont {Y.}~\bibnamefont {Zhao}},\ }\href {\doibase
  https://doi.org/10.1002/andp.201800120} {\bibfield  {journal} {\bibinfo
  {journal} {Ann. Phys. (Berlin)}\ }\textbf {\bibinfo {volume} {530}},\
  \bibinfo {pages} {1800120} (\bibinfo {year} {2018})}\BibitemShut {NoStop}%
\bibitem [{\citenamefont {Nägele}\ and\ \citenamefont
  {Weiss}(2010)}]{Nagele10}%
  \BibitemOpen
  \bibfield  {author} {\bibinfo {author} {\bibfnamefont {P.}~\bibnamefont
  {Nägele}}\ and\ \bibinfo {author} {\bibfnamefont {U.}~\bibnamefont
  {Weiss}},\ }\href {\doibase https://doi.org/10.1016/j.physe.2009.06.060}
  {\bibfield  {journal} {\bibinfo  {journal} {Physica E: Low-dim. Syst.
  Nanostruct.}\ }\textbf {\bibinfo {volume} {42}},\ \bibinfo {pages} {622}
  (\bibinfo {year} {2010})},\ \bibinfo {note} {proceedings of the international
  conference Frontiers of Quantum and Mesoscopic Thermodynamics FQMT
  '08}\BibitemShut {NoStop}%
\bibitem [{\citenamefont {Storcz}\ \emph {et~al.}(2005)\citenamefont {Storcz},
  \citenamefont {Hellmann}, \citenamefont {Hrelescu},\ and\ \citenamefont
  {Wilhelm}}]{Storcz05}%
  \BibitemOpen
  \bibfield  {author} {\bibinfo {author} {\bibfnamefont {M.~J.}\ \bibnamefont
  {Storcz}}, \bibinfo {author} {\bibfnamefont {F.}~\bibnamefont {Hellmann}},
  \bibinfo {author} {\bibfnamefont {C.}~\bibnamefont {Hrelescu}}, \ and\
  \bibinfo {author} {\bibfnamefont {F.~K.}\ \bibnamefont {Wilhelm}},\ }\href
  {\doibase 10.1103/PhysRevA.72.052314} {\bibfield  {journal} {\bibinfo
  {journal} {Phys. Rev. A}\ }\textbf {\bibinfo {volume} {72}},\ \bibinfo
  {pages} {052314} (\bibinfo {year} {2005})}\BibitemShut {NoStop}%
\bibitem [{\citenamefont {Garst}\ \emph {et~al.}(2004)\citenamefont {Garst},
  \citenamefont {Kehrein}, \citenamefont {Pruschke}, \citenamefont {Rosch},\
  and\ \citenamefont {Vojta}}]{Garst04}%
  \BibitemOpen
  \bibfield  {author} {\bibinfo {author} {\bibfnamefont {M.}~\bibnamefont
  {Garst}}, \bibinfo {author} {\bibfnamefont {S.}~\bibnamefont {Kehrein}},
  \bibinfo {author} {\bibfnamefont {T.}~\bibnamefont {Pruschke}}, \bibinfo
  {author} {\bibfnamefont {A.}~\bibnamefont {Rosch}}, \ and\ \bibinfo {author}
  {\bibfnamefont {M.}~\bibnamefont {Vojta}},\ }\href {\doibase
  10.1103/PhysRevB.69.214413} {\bibfield  {journal} {\bibinfo  {journal} {Phys.
  Rev. B}\ }\textbf {\bibinfo {volume} {69}},\ \bibinfo {pages} {214413}
  (\bibinfo {year} {2004})}\BibitemShut {NoStop}%
\bibitem [{\citenamefont {McCutcheon}\ \emph {et~al.}(2010)\citenamefont
  {McCutcheon}, \citenamefont {Nazir}, \citenamefont {Bose},\ and\
  \citenamefont {Fisher}}]{McCutcheon10}%
  \BibitemOpen
  \bibfield  {author} {\bibinfo {author} {\bibfnamefont {D.~P.~S.}\
  \bibnamefont {McCutcheon}}, \bibinfo {author} {\bibfnamefont
  {A.}~\bibnamefont {Nazir}}, \bibinfo {author} {\bibfnamefont
  {S.}~\bibnamefont {Bose}}, \ and\ \bibinfo {author} {\bibfnamefont {A.~J.}\
  \bibnamefont {Fisher}},\ }\href {\doibase 10.1103/PhysRevB.81.235321}
  {\bibfield  {journal} {\bibinfo  {journal} {Phys. Rev. B}\ }\textbf {\bibinfo
  {volume} {81}},\ \bibinfo {pages} {235321} (\bibinfo {year}
  {2010})}\BibitemShut {NoStop}%
\bibitem [{\citenamefont {Bonart}(2013)}]{Bonart13}%
  \BibitemOpen
  \bibfield  {author} {\bibinfo {author} {\bibfnamefont {J.}~\bibnamefont
  {Bonart}},\ }\href {\doibase 10.1103/PhysRevB.88.125139} {\bibfield
  {journal} {\bibinfo  {journal} {Phys. Rev. B}\ }\textbf {\bibinfo {volume}
  {88}},\ \bibinfo {pages} {125139} (\bibinfo {year} {2013})}\BibitemShut
  {NoStop}%
\bibitem [{\citenamefont {Orth}\ \emph {et~al.}(2010)\citenamefont {Orth},
  \citenamefont {Roosen}, \citenamefont {Hofstetter},\ and\ \citenamefont
  {Le~Hur}}]{Orth10}%
  \BibitemOpen
  \bibfield  {author} {\bibinfo {author} {\bibfnamefont {P.~P.}\ \bibnamefont
  {Orth}}, \bibinfo {author} {\bibfnamefont {D.}~\bibnamefont {Roosen}},
  \bibinfo {author} {\bibfnamefont {W.}~\bibnamefont {Hofstetter}}, \ and\
  \bibinfo {author} {\bibfnamefont {K.}~\bibnamefont {Le~Hur}},\ }\href
  {\doibase 10.1103/PhysRevB.82.144423} {\bibfield  {journal} {\bibinfo
  {journal} {Phys. Rev. B}\ }\textbf {\bibinfo {volume} {82}},\ \bibinfo
  {pages} {144423} (\bibinfo {year} {2010})}\BibitemShut {NoStop}%
\bibitem [{\citenamefont {Zheng}\ \emph {et~al.}(2015)\citenamefont {Zheng},
  \citenamefont {L\"u},\ and\ \citenamefont {Zhao}}]{Zheng15}%
  \BibitemOpen
  \bibfield  {author} {\bibinfo {author} {\bibfnamefont {H.}~\bibnamefont
  {Zheng}}, \bibinfo {author} {\bibfnamefont {Z.}~\bibnamefont {L\"u}}, \ and\
  \bibinfo {author} {\bibfnamefont {Y.}~\bibnamefont {Zhao}},\ }\href {\doibase
  10.1103/PhysRevE.91.062115} {\bibfield  {journal} {\bibinfo  {journal} {Phys.
  Rev. E}\ }\textbf {\bibinfo {volume} {91}},\ \bibinfo {pages} {062115}
  (\bibinfo {year} {2015})}\BibitemShut {NoStop}%
\bibitem [{\citenamefont {Winter}\ and\ \citenamefont
  {Rieger}(2014)}]{Winter14}%
  \BibitemOpen
  \bibfield  {author} {\bibinfo {author} {\bibfnamefont {A.}~\bibnamefont
  {Winter}}\ and\ \bibinfo {author} {\bibfnamefont {H.}~\bibnamefont
  {Rieger}},\ }\href {\doibase 10.1103/PhysRevB.90.224401} {\bibfield
  {journal} {\bibinfo  {journal} {Phys. Rev. B}\ }\textbf {\bibinfo {volume}
  {90}},\ \bibinfo {pages} {224401} (\bibinfo {year} {2014})}\BibitemShut
  {NoStop}%
\bibitem [{\citenamefont {Nägele}\ \emph {et~al.}(2008)\citenamefont
  {Nägele}, \citenamefont {Campagnano},\ and\ \citenamefont
  {Weiss}}]{Nagele08}%
  \BibitemOpen
  \bibfield  {author} {\bibinfo {author} {\bibfnamefont {P.}~\bibnamefont
  {Nägele}}, \bibinfo {author} {\bibfnamefont {G.}~\bibnamefont {Campagnano}},
  \ and\ \bibinfo {author} {\bibfnamefont {U.}~\bibnamefont {Weiss}},\ }\href
  {\doibase 10.1088/1367-2630/10/11/115010} {\bibfield  {journal} {\bibinfo
  {journal} {New J. Phys.}\ }\textbf {\bibinfo {volume} {10}},\ \bibinfo
  {pages} {115010} (\bibinfo {year} {2008})}\BibitemShut {NoStop}%
\bibitem [{\citenamefont {Thorwart}\ and\ \citenamefont
  {H\"anggi}(2001)}]{Thorwart01}%
  \BibitemOpen
  \bibfield  {author} {\bibinfo {author} {\bibfnamefont {M.}~\bibnamefont
  {Thorwart}}\ and\ \bibinfo {author} {\bibfnamefont {P.}~\bibnamefont
  {H\"anggi}},\ }\href {\doibase 10.1103/PhysRevA.65.012309} {\bibfield
  {journal} {\bibinfo  {journal} {Phys. Rev. A}\ }\textbf {\bibinfo {volume}
  {65}},\ \bibinfo {pages} {012309} (\bibinfo {year} {2001})}\BibitemShut
  {NoStop}%
\bibitem [{\citenamefont {Storcz}\ and\ \citenamefont
  {Wilhelm}(2003)}]{Storcz03}%
  \BibitemOpen
  \bibfield  {author} {\bibinfo {author} {\bibfnamefont {M.~J.}\ \bibnamefont
  {Storcz}}\ and\ \bibinfo {author} {\bibfnamefont {F.~K.}\ \bibnamefont
  {Wilhelm}},\ }\href {\doibase 10.1103/PhysRevA.67.042319} {\bibfield
  {journal} {\bibinfo  {journal} {Phys. Rev. A}\ }\textbf {\bibinfo {volume}
  {67}},\ \bibinfo {pages} {042319} (\bibinfo {year} {2003})}\BibitemShut
  {NoStop}%
\bibitem [{\citenamefont {Calvo}\ \emph {et~al.}(2011)\citenamefont {Calvo},
  \citenamefont {Abud}, \citenamefont {Sartoris},\ and\ \citenamefont
  {Santana}}]{Calvo11}%
  \BibitemOpen
  \bibfield  {author} {\bibinfo {author} {\bibfnamefont {R.}~\bibnamefont
  {Calvo}}, \bibinfo {author} {\bibfnamefont {J.~E.}\ \bibnamefont {Abud}},
  \bibinfo {author} {\bibfnamefont {R.~P.}\ \bibnamefont {Sartoris}}, \ and\
  \bibinfo {author} {\bibfnamefont {R.~C.}\ \bibnamefont {Santana}},\ }\href
  {\doibase 10.1103/PhysRevB.84.104433} {\bibfield  {journal} {\bibinfo
  {journal} {Phys. Rev. B}\ }\textbf {\bibinfo {volume} {84}},\ \bibinfo
  {pages} {104433} (\bibinfo {year} {2011})}\BibitemShut {NoStop}%
\bibitem [{\citenamefont {Napolitano}\ \emph {et~al.}(2008)\citenamefont
  {Napolitano}, \citenamefont {Nascimento}, \citenamefont {Cabaleiro},
  \citenamefont {Castro},\ and\ \citenamefont {Calvo}}]{Napolitano08}%
  \BibitemOpen
  \bibfield  {author} {\bibinfo {author} {\bibfnamefont {L.~M.~B.}\
  \bibnamefont {Napolitano}}, \bibinfo {author} {\bibfnamefont {O.~R.}\
  \bibnamefont {Nascimento}}, \bibinfo {author} {\bibfnamefont
  {S.}~\bibnamefont {Cabaleiro}}, \bibinfo {author} {\bibfnamefont
  {J.}~\bibnamefont {Castro}}, \ and\ \bibinfo {author} {\bibfnamefont
  {R.}~\bibnamefont {Calvo}},\ }\href {\doibase 10.1103/PhysRevB.77.214423}
  {\bibfield  {journal} {\bibinfo  {journal} {Phys. Rev. B}\ }\textbf {\bibinfo
  {volume} {77}},\ \bibinfo {pages} {214423} (\bibinfo {year}
  {2008})}\BibitemShut {NoStop}%
\bibitem [{\citenamefont {Vandersypen}\ and\ \citenamefont
  {Chuang}(2005)}]{Vandersypen05}%
  \BibitemOpen
  \bibfield  {author} {\bibinfo {author} {\bibfnamefont {L.~M.~K.}\
  \bibnamefont {Vandersypen}}\ and\ \bibinfo {author} {\bibfnamefont {I.~L.}\
  \bibnamefont {Chuang}},\ }\href {\doibase 10.1103/RevModPhys.76.1037}
  {\bibfield  {journal} {\bibinfo  {journal} {Rev. Mod. Phys.}\ }\textbf
  {\bibinfo {volume} {76}},\ \bibinfo {pages} {1037} (\bibinfo {year}
  {2005})}\BibitemShut {NoStop}%
\bibitem [{\citenamefont {Weidt}\ \emph {et~al.}(2016)\citenamefont {Weidt},
  \citenamefont {Randall}, \citenamefont {Webster}, \citenamefont {Lake},
  \citenamefont {Webb}, \citenamefont {Cohen}, \citenamefont {Navickas},
  \citenamefont {Lekitsch}, \citenamefont {Retzker},\ and\ \citenamefont
  {Hensinger}}]{Weidt16}%
  \BibitemOpen
  \bibfield  {author} {\bibinfo {author} {\bibfnamefont {S.}~\bibnamefont
  {Weidt}}, \bibinfo {author} {\bibfnamefont {J.}~\bibnamefont {Randall}},
  \bibinfo {author} {\bibfnamefont {S.~C.}\ \bibnamefont {Webster}}, \bibinfo
  {author} {\bibfnamefont {K.}~\bibnamefont {Lake}}, \bibinfo {author}
  {\bibfnamefont {A.~E.}\ \bibnamefont {Webb}}, \bibinfo {author}
  {\bibfnamefont {I.}~\bibnamefont {Cohen}}, \bibinfo {author} {\bibfnamefont
  {T.}~\bibnamefont {Navickas}}, \bibinfo {author} {\bibfnamefont
  {B.}~\bibnamefont {Lekitsch}}, \bibinfo {author} {\bibfnamefont
  {A.}~\bibnamefont {Retzker}}, \ and\ \bibinfo {author} {\bibfnamefont
  {W.~K.}\ \bibnamefont {Hensinger}},\ }\href {\doibase
  10.1103/PhysRevLett.117.220501} {\bibfield  {journal} {\bibinfo  {journal}
  {Phys. Rev. Lett.}\ }\textbf {\bibinfo {volume} {117}},\ \bibinfo {pages}
  {220501} (\bibinfo {year} {2016})}\BibitemShut {NoStop}%
\bibitem [{\citenamefont {Gaetan}\ \emph {et~al.}(2009)\citenamefont {Gaetan},
  \citenamefont {Miroshnychenko}, \citenamefont {Wilk}, \citenamefont {Chotia},
  \citenamefont {Viteau}, \citenamefont {Comparat}, \citenamefont {Pillet},
  \citenamefont {Browaeys},\ and\ \citenamefont {Grangier}}]{Gaetan09}%
  \BibitemOpen
  \bibfield  {author} {\bibinfo {author} {\bibfnamefont {A.}~\bibnamefont
  {Gaetan}}, \bibinfo {author} {\bibfnamefont {Y.}~\bibnamefont
  {Miroshnychenko}}, \bibinfo {author} {\bibfnamefont {T.}~\bibnamefont
  {Wilk}}, \bibinfo {author} {\bibfnamefont {A.}~\bibnamefont {Chotia}},
  \bibinfo {author} {\bibfnamefont {M.}~\bibnamefont {Viteau}}, \bibinfo
  {author} {\bibfnamefont {D.}~\bibnamefont {Comparat}}, \bibinfo {author}
  {\bibfnamefont {P.}~\bibnamefont {Pillet}}, \bibinfo {author} {\bibfnamefont
  {A.}~\bibnamefont {Browaeys}}, \ and\ \bibinfo {author} {\bibfnamefont
  {P.}~\bibnamefont {Grangier}},\ }\href@noop {} {\bibfield  {journal}
  {\bibinfo  {journal} {Nat. Phys.}\ }\textbf {\bibinfo {volume} {5}},\
  \bibinfo {pages} {115} (\bibinfo {year} {2009})}\BibitemShut {NoStop}%
\bibitem [{\citenamefont {Urban}\ \emph {et~al.}(2009)\citenamefont {Urban},
  \citenamefont {Johnson}, \citenamefont {Henage}, \citenamefont {Isenhower},
  \citenamefont {Yavuz}, \citenamefont {Walker},\ and\ \citenamefont
  {Saffman}}]{Urban09}%
  \BibitemOpen
  \bibfield  {author} {\bibinfo {author} {\bibfnamefont {E.}~\bibnamefont
  {Urban}}, \bibinfo {author} {\bibfnamefont {T.~A.}\ \bibnamefont {Johnson}},
  \bibinfo {author} {\bibfnamefont {T.}~\bibnamefont {Henage}}, \bibinfo
  {author} {\bibfnamefont {L.}~\bibnamefont {Isenhower}}, \bibinfo {author}
  {\bibfnamefont {D.}~\bibnamefont {Yavuz}}, \bibinfo {author} {\bibfnamefont
  {T.}~\bibnamefont {Walker}}, \ and\ \bibinfo {author} {\bibfnamefont
  {M.}~\bibnamefont {Saffman}},\ }\href@noop {} {\bibfield  {journal} {\bibinfo
   {journal} {Nat. Phys.}\ }\textbf {\bibinfo {volume} {5}},\ \bibinfo {pages}
  {110} (\bibinfo {year} {2009})}\BibitemShut {NoStop}%
\end{thebibliography}%

\end{document}